\documentclass[useAMS,usenatbib,usegraphicx]{mn2e}

\title[Frequency-dependent core shifts in AGN jets.]{A new method
for estimating frequency-dependent core shifts in AGN jets.}
\author[Kudryavtseva et al.]{N.A.~Kudryavtseva,$^{1}$\thanks{E-mail: nadia.kudryavtseva@curtin.edu.au}
D.C.~Gabuzda,$^{2}$ M.F.~Aller$^{3}$ and H.D.~Aller$^{3}$\\
$^{1}$ Curtin Institute of Radio Astronomy, Curtin University, GPO Box U1987, Perth, WA 6845, Australia \\
$^{2}$Physics Department, University College Cork, Cork, Ireland \\
$^{3}$ Astronomy Department, University of Michigan, Ann Arbor, MI 48109-1042, USA\\
}
\begin{document}

%\date{Accepted 1988 December 15. Received 1988 December 14; in original form 1988 October 11}

\pagerange{\pageref{firstpage}--\pageref{lastpage}} \pubyear{2010}

\maketitle

\label{firstpage}

\begin{abstract}
We discuss the opacity in the core regions of Active Galactic Nuclei (AGN) observed with Very Long Baseline radio Interferometry (VLBI) and
describe a new method for deriving the frequency-dependent shifts of the VLBI core from the frequency-dependent time lags of flares observed
with single-dish observations. Application of the method to the core shifts of the quasar 3C~345 shows a very good agreement between the core
shifts directly measured from VLBI observations and derived from flares in the the total flux-density using the proposed method. This provides
direct evidence that the observed jet component speeds in this AGN represent the actual physical speed of the feature, rather than a pattern
speed. The frequency-dependent time lags of flares can be used to derive physical parameters of the jets, such as distance from the VLBI core
to the base of the jet and the magnetic fields in the core region. Our estimates for 3C~345 indicate core magnetic fields $\simeq 0.1$~G and
magnetic field at 1~pc $\simeq0.4$~G.
\end{abstract}


\section{Introduction}

Most Active Galactic Nuclei (AGN) display a ``core+jet'' structure in Very Long Baseline Interferometry (VLBI) images. In the standard
interpretation, the core is taken to be the optically thick base of the jet~\citep{BK_1979}. Due to synchrotron self-absorption, the absolute
position of the observed VLBI core ($\tau=1$ surface) shifts systematically with frequency, moving increasingly outward along the VLBI jet with
lower frequency~\citep{Konigl_1981}.  This frequency-dependent core shift has a direct effect on astrometric measurements performed in the
radio and optical. In the near future, the GAIA astrometry mission and the Space Interferometry Mission will begin. For both missions, matching
the optical astrometric catalogues to the radio catalogues~\citep[e.g.][]{Fey_2001} presents a very important problem. The core shifts can
introduce offsets between the optical and radio positions of AGN of up to several milliarcseconds~\citep{Kovalev_2008}, which will strongly
affect the accuracy of matching the radio and optical catalogues.

Core shifts are also needed for the correct reconstruction of VLBI
spectral-index and rotation-measure maps. Moreover, knowledge of the
frequency-dependent core shifts can be used to derive physical parameters of
the jet, such as the core magnetic field and the distance from the VLBI core
to the base of the jet~\citep{Lobanov_1998,Hirotani_2005}.

Therefore, precise measurements of the core shifts are necessary. One way
to obtain core shifts is through phase-referencing VLBI observations,
however this is a complex and resource-intensive technique, and
phase-referencing core shifts have been determined for only a few AGN,
such as 1038+528~\citep{Marcaide_1984}, 4C~39.25~\citep{Guirado_1995},
3C~395~\citep{Lara_1994}, 3C~390.1~\citep{Ros_2001}, and
M~81~\citep{Bietenholz_2004}. Another indirect method to measure core
shifts is to align optically thin parts of an AGN jet at different
frequencies~\citep[e.g.][]{Kovalev_2008,Croke_2008,Osullivan_2009}. However,
this method requires simultaneous multi-frequency VLBI observations,
which are likewise fairly resource intensive, and does not always yield
unambiguous results. The limitations of these techniques are exacerbated
by the fact that the core shift may well depend on the activity state of
an AGN, and therefore be time dependent, whereas at present only isolated
core-shift measurements for individual AGN are available. In addition, the
magnitude of the core shifts that can be detected is limited by the
resolution of the VLBI observations used.

In this paper, we discuss the opacity in the core regions of AGN and describe
a new method for deriving core shifts from the frequency-dependent time
lags of flares observed with single-dish observations. We use these time
lags to derive physical parameters of the jets, such as the distance
from the VLBI core to the base of the jet and the magnetic fields in the
core region.

\section{Frequency-dependent time lags}
\label{section_time_lags}

\begin{figure*}
\centering
\includegraphics[clip,width=14cm]{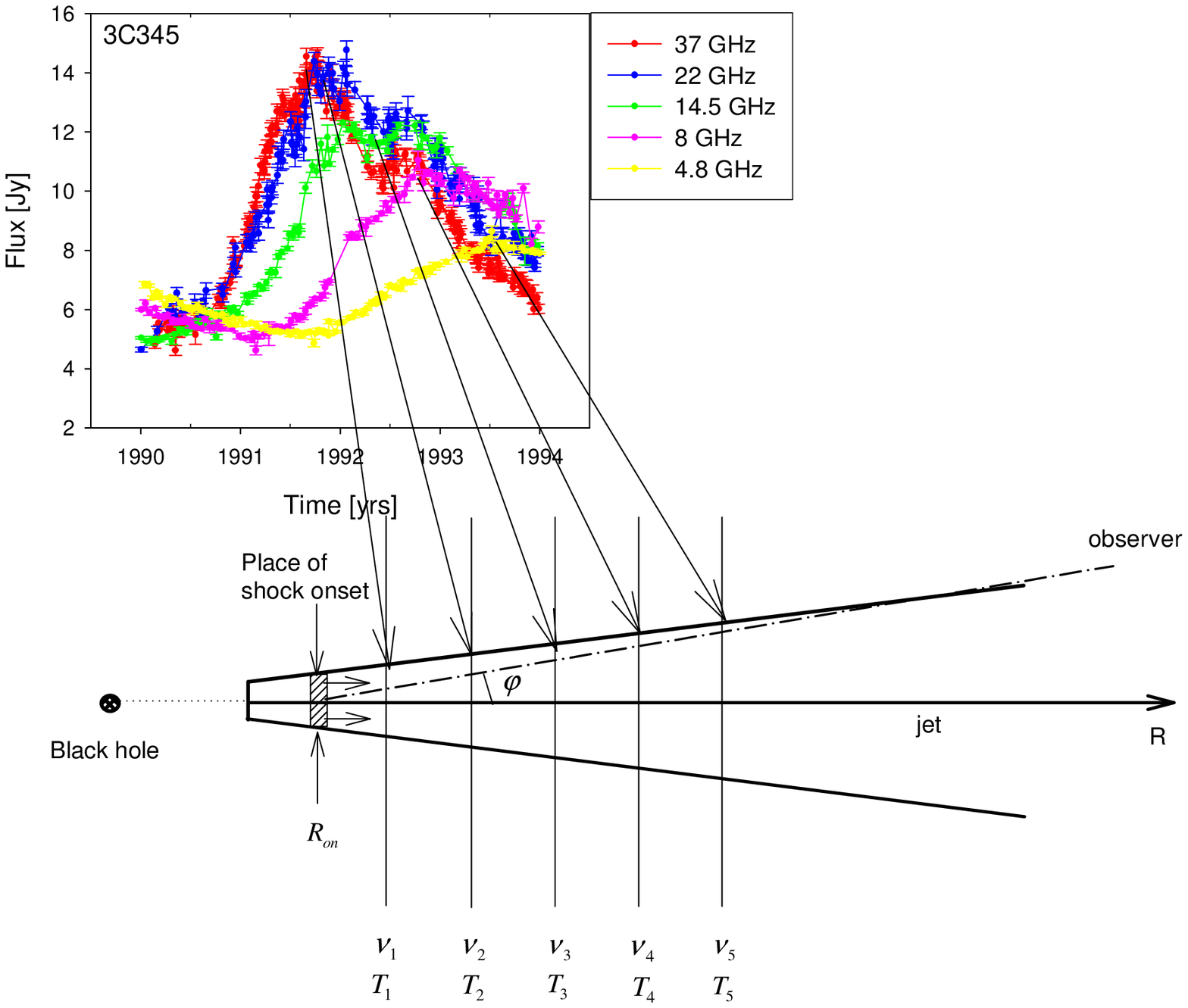} \caption[]{Sketch of the jet geometry discussed in section~\ref{section_time_lags} as observed
in the rest frame of the quasar. The shaded rectangle shows a shock wave
that starts at the distance $R_{on}$ and propagates along the jet axis with
the speed $\beta$. The vertical lines mark the core positions at different
frequencies and the times when the shock wave reaches the core at a
particular frequency. The times $T_{i}$ correspond to the maxima of the flares
at frequencies $\nu_{i}$.} \label{jet}
\end{figure*}

The core is a compact feature in the VLBI map of an AGN with high brightness
and a relatively flat spectrum, which is usually interpreted as that part
of the jet
where the optical depth is $\tau = 1$. Since the $\tau = 1$ surface has
different locations at different frequencies~\citep{Konigl_1981}, the absolute
position of the core should depend on the observing frequency. According to
the standard shock-in-jet model, a change of electron density and pressure
at the injection point near the base of the jet will cause a shock wave
to propagate along the jet. This will cause brightening of the core region,
which will be seen as a flare in the total flux-density light curves, and will
be followed by the appearance of a jet component (or components) in VLBI
maps~\citep[e.g.][]{Marscher_Gear_1985,Gomez_1997}. Consider a conical jet
geometry (see Figure~\ref{jet}) observed at a viewing angle
$\varphi$. A shock wave appears at the distance $R_{on}$ and then propagates
along the jet. It will cross the $\tau = 1$ surface for a particular frequency
$\nu_{i}$ at some time $T_{i}$ and emerge out of the core region.
Since the position of the $\tau = 1$ surface is shifted further from the
jet base at lower frequencies, the times $T_{i}$ will be delayed at lower
frequencies.
Figure~\ref{jet} shows how the $T_{i}$ correspond to the maxima of the total
flux-density flares obtained with single-dish observations. Crossing the
$\tau = 1$ surface corresponds to a maximum of the total flux-density
outburst at the corresponding frequency, and the time of the flare maximum
thus also depends on the frequency; thus, the time lags contain information
about core opacity. According to the geometry of Figure~\ref{jet}
and the principles of superluminal
motion~\citep[e.g.][]{Rees_1967,Tuerler_2000}, the time $t$ that has passed
after the shock wave's appearance at the distance $R_{on}$ in the observer's
frame is
\begin{equation}
t = \frac{(1 + z) sin\varphi(R(\nu) - R_{on}(\nu))}{\beta_{app} c},
\end{equation}
where $R$ is the distance along the jet axis in the rest frame of the quasar,
$\beta_{app}$ is the apparent velocity in the plane of the sky in units of
$c$, $\varphi$ is the viewing angle, and $z$ is the redshift of the source.
In this equation, we assume that the apparent speed is the actual speed
of the shock. Observing at multiple frequencies $\nu_{a}$ and $\nu_{b}$,
we can estimate the time lags between the maxima of the total flux-density
outburst:
%and write equation for R at the distance where optical depth reaches unity
\begin{eqnarray}
\Delta t & = & t(\nu_{a}) - t(\nu_{b})\\
   & = &\frac{(1+z) sin\varphi}{\beta_{app} c} [R(\nu_{a}) - R(\nu_{b}) +
R_{on}(\nu_{a}) - R_{on}(\nu_{b})].
\end{eqnarray}
Assuming that the shock wave appears at the same distance $R_{on}$ at
all frequencies, we obtain $\Delta
t(\nu)_{obs} = \Delta R_{proj}/{\beta_{app} c}$, where $\Delta R_{proj}$ is
the distance between $R$ at the two frequencies projected onto the plane
of the sky. Taking expression (33) for the frequency-dependent core shift
from~\citet{Hirotani_2005}, we can write the dependence of the time lags
on the frequency, spectral index $\alpha$, magnetic field and electron
density:
\begin{equation}
\Delta t(\nu)_{obs} = \frac{sin\varphi}{\beta_{app}} (x_{B}^{k_{b}} f \nu_{0})^{1/k_{r}} \frac{\nu_{b}^{1/k_{r}} - \nu_{a}^{1/k_{r}}
}{\nu_{a}^{1/k_{r}} \nu_{b}^{1/k_{r}}}, \label{eq:dt}
\end{equation}
where
\begin{equation}
k_{r} \equiv \frac{(3 - 2\alpha)m + 2n - 2}{5 - 2\alpha}, \label{k_r}
\end{equation}
\begin{equation}
k_{B} \equiv \frac{3 - 2\alpha}{5 - 2\alpha},
\end{equation}
$x_{B}$ is a dimensionless variable for $B$ and $f$ is a function of $N_{1}$,
the spectral index, and the viewing angle.

Defining the core-position (or time-lag) offset as
\begin{equation} \label{eq:omega}
\Omega_{r\nu} = 4.85 \times 10^{-9} \frac{\Delta t \beta_{app} D_{L}}{(1+z)^2} \left(\frac{\nu_{a}^{1/k_{r}}
\nu_{b}^{1/k_{r}}}{\nu_{b}^{1/k_{r}} - \nu_{a}^{1/k_{r}}}\right)
\end{equation}
we can obtain formulas for the magnetic field and the distance from the VLBI
core to the central engine. In this formula, $\Omega_{r\nu}$ is measured
in units of pc~$\cdot$~GHz, $\Delta t$ in yrs, $\beta_{app}$ in milliarcsecond
(mas) per year, and $D_{L}$ is the luminosity distance of the source in
parsec.

\begin{figure*}
\centering
\includegraphics[clip,width=14cm]{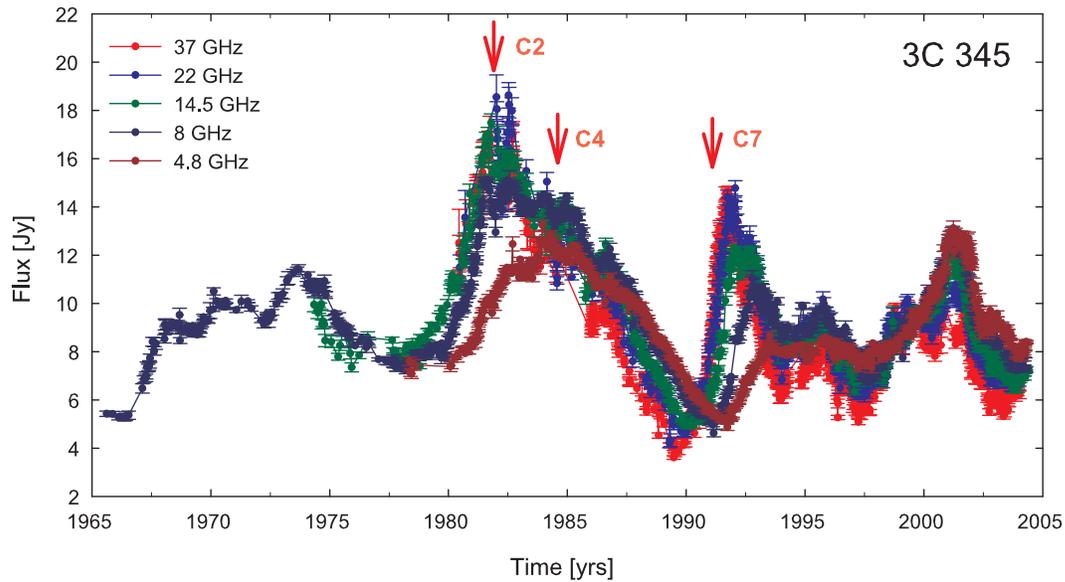} \caption[Total
flux-density light curve of 3C~345]{Total flux-density light
curve of 3C~345 at 4.8~GHz, 8~GHz, 14.5~GHz, 22~GHz, and 37~GHz.
The data are from the University of Michigan and Mets\"{a}hovi Radio
Observatory monitoring databases. The arrows mark the times of the
ejections of the new jet components C2, C4 and C7.}
\label{3c345_lcurve}
\end{figure*}

Formula \ref{eq:dt} assumes that the electron number density and
magnetic field scale with distance along the jet $R$ as
\begin{equation}
N_{e}^* = N_{1} r^{-n}, B = B_{1} r^{-m}, \label{nm}
\end{equation}
where $N_{1}$ and $B_{1}$ refer to the values of $N_{e}^*$ and $B$ at
$r = 1$~pc.

Following~\citet{Lobanov_1998} and~\citet{Hirotani_2005}, we can write the
distance between the core and the base of the jet in terms of the
frequency-dependent time lags as
\begin{equation} \label{eq:rcore}
r_{core}(\nu) = \frac{\Omega_{r\nu}}{sin \varphi} \nu^{-1/k_{r}},
\end{equation}
where $k_r$ is estimated using the frequency-dependent time lags. We can calculate the magnetic field at 1~pc distance for the special case
when there is equipartition between the energies of the particles and magnetic field ($k_{r} = 1$) and the spectral index is $\alpha =
-0.5$~\citep{Osullivan_2009}:
\begin{equation} \label{eq:b1}
B_{1} \cong 0.025 \left(\frac{\Omega_{r\nu}^3 (1+z)^2}{\delta^2 \theta \sin^2 \varphi}\right)^{1/4},
\end{equation}
where $\delta$ is the Doppler factor, $\theta$ the jet half-opening angle,
$\varphi$ the viewing angle, and $B_{1}$ is in Gauss. The equipartition magnetic-field
strength in the core can then be found from the relation
\begin{equation} \label{eq:bcore}
B_{core}(\nu) = B_{1} r_{core}^{-1}
\end{equation}
for $B_{core}$ at a particular frequency $\nu$.

\section{Fitting the flux-density light curves}
In order to check the proposed method for measuring the frequency-dependent core shifts from the time lags and the validity of our assumptions,
we calculated time lags from the total flux-density light curves for the AGN 3C~345, whose core shift has been measured at the same frequencies
used to construct the light curves. The frequency-dependent time delays were calculated by fitting Gaussian functions to the total flux-density
light curves from the University of Michigan~\citep{Aller_1999,Aller_2003} and Mets\"{a}hovi Radio
Observatory~\citep{terasranta_1992,terasranta_1998,terasranta_2004, terasranta_2005} monitoring databases at 4.8~GHz, 8~GHz, 14.5~GHz, 22~GHz,
and 37~GHz.

The total flux-density light curves were decomposed into Gaussian components
following the procedure discussed in~\citet{Pyatunina_2006}
and~\citet{Pyatunina_2007}. Long-term trends in the total flux-density
variations were calculated as polynomial approximations for the deepest
minima in the light curves and subtracted before the fitting. The Gaussian
decomposition was performed such that it first removed the trend, then
found the highest peak in the light curve and fitted a Gaussian to the peak,
based on a $\chi^2$ minimization. The fitted component was then removed
from the light curve and the procedure was repeated until all significant
peaks were fitted with Gaussians. During the fitting, we applied the general
criterion that the smallest number of individual flares (Gaussian components)
providing a complete description of the light curve was used. The number of
fitted Gaussians depended on the time interval covered by the light curve and
the characteristic time scale of the source variability.

\begin{figure}
\centering
\includegraphics[clip,width=9cm]{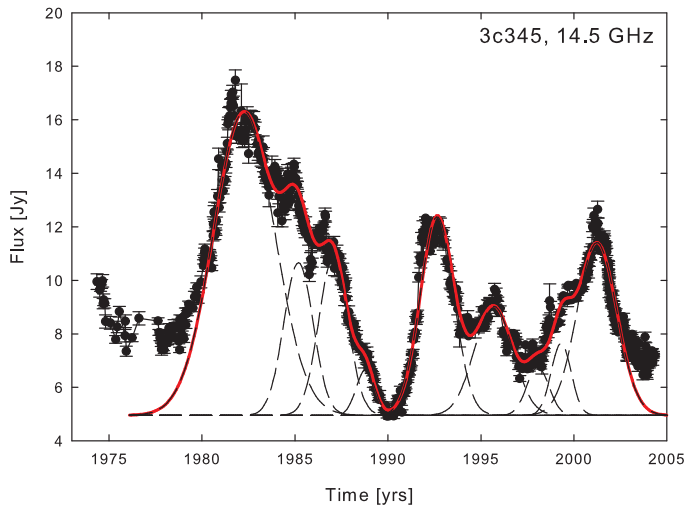} \caption[]{Total
flux-density light curve of 3C~345 at 14.5~GHz (the data are from the
University of Michigan monitoring database). The long dashed lines show
decomposed Gaussian components and the solid line the sum of
the fitted Gaussians. } \label{3c345_decomp}
\end{figure}

\section{Test of the Method: The Quasar 3C~345}

The quasar 3C~345 (z = 0.5928,~\citet{Marziani_1996}) is one of the best
studied AGN on VLBI scales. Several jet components with apparent velocities
of $2-20c$ have been observed, moving along strongly curved
trajectories~\citep*[e.g.][]{Unwin_1983,Zensus_1995,Lobanov_1999,
Rantakyro_1995,Ros_2000}. \citet{Klare_2005} found evidence for periodic
changes in the parsec-scale jet with a 9-year period.

The total flux-density light curves at 4.8~GHz, 8~GHz, 14.5~GHz, 22~GHz and 37~GHz are shown in Fig.~\ref{3c345_lcurve}.  The best $\chi^2$
values were reached with a fit of 9 Gaussians to the light curve; the Gaussians fitted are shown together with the light curve at 14.5~GHz in
Fig.~\ref{3c345_decomp}. The long-dashed curves show the Gaussians and the solid curve the final sum of all 9 fitted Gaussians. It is clear
that the sum of the Gaussians fits well all the main features in the light curve.  The parameters of the 9 Gaussians are given in
Table~\ref{3c345_flares}. The columns of this table give the (1) component designation, (2) observing frequency, (3) epoch of maximum flux, (4)
maximum amplitude, (5) full width at half maximum (FWHM) of the Gaussian component corresponding to the outburst $\Theta$, (6) time delay
between the epoch of maximum flux for the outbursts at the given frequency and the corresponding epoch at the highest available frequency
$\Delta T$, (7) calculated spectral indexes and (8) calculated $k_{r}$ values (the $k_r$ values will be discussed below). We use a positive
spectral index convention $S \sim \nu^\alpha$. We measured the spectral index for each flare, fitting a linear regression into log($\nu$) vs.
log(S) plots, where S is an amplitude of a flare, obtained as the maximum of fitted Gaussian function. Spectral index was calculated only for
those outbursts that have been detected at three or more frequencies.

We were able to calculate time-delay core shifts for the epochs of three
outbursts associated with the ejection of components for which superluminal
speeds have been published: C2, C4 and C7, marked in Fig.~\ref{3c345_lcurve}.
%Using what equation? What was phi?
Core-shift measurements based on a direct comparison of VLBI images
at different frequencies are available for epochs near our measured time
delays.  Table~\ref{3C345_shifts} summarizes and compares the core shifts
directly measured from the VLBI images and our calculated core shifts
derived from time lags obtained with single dish observations. The epoch of
the VLBI observations and name of the ejected jet components used for the
analysis are listed in this table as well.

\begin{table*}
\caption{3C~345: Parameters of outbursts} \centering
\begin{tabular}{lrllllcc}
\hline
Comp.& Freq. & Amplitude     & $T_{max}$  & $\Theta$ & Time delay  & Sp.Index & $k_{r}$ \\
      & (GHz) & (Jy)        & (yr)       &   (yr)   &  (yr)        &          &         \\
\hline

$A$ & 37.0  & $12.69\pm0.10$ & $1982.19\pm0.02$ &  $3.90\pm0.05$ & $0.00\pm0.04$ & $0.30\pm0.07$ & $1.75\pm0.02$ \\
    & 22.0  & $11.79\pm0.06$ & $1982.29\pm0.07$ &  $6.05\pm0.08$ & $0.11\pm0.09$ &               &               \\
    & 14.5  & $11.35\pm0.04$ & $1982.27\pm0.01$ &  $3.87\pm0.02$ & $0.09\pm0.03$ &               &               \\
    & 8.0   & $9.49\pm0.01$  & $1982.74\pm0.01$ &  $4.81\pm0.01$ & $0.55\pm0.03$ &               &               \\
    & 4.8   & $6.56\pm0.03$  & $1982.86\pm0.03$ &  $3.92\pm0.03$ & $0.67\pm0.05$ &               &               \\
$B$ & 14.5  & $5.71\pm0.03$  & $1985.19\pm0.01$ &  $1.86\pm0.01$ & $0.00\pm0.03$ & $0.98\pm0.79$ & --            \\
    & 8.0   & $1.52\pm0.02$  & $1985.05\pm0.01$ &  $1.25\pm0.01$ & $0.14\pm0.02$ &               &               \\
    & 4.8   & $2.01\pm0.02$  & $1985.57\pm0.01$ &  $1.18\pm0.01$ & $0.38\pm0.02$ &               &               \\
$C$ & 37.0  & $5.54\pm0.04$  & $1986.58\pm0.01$ &  $2.53\pm0.03$ & $0.00\pm0.02$ & --            & $1.81\pm0.05$ \\
    & 22.0  & $3.45\pm0.04$  & $1987.05\pm0.04$ &  $3.38\pm0.04$ & $0.47\pm0.05$ &               &               \\
    & 14.5  & $5.88\pm0.03$  & $1987.07\pm0.01$ &  $1.84\pm0.01$ & $0.49\pm0.02$ &               &               \\
    & 8.0   & $5.12\pm0.02$  & $1987.09\pm0.01$ &  $3.62\pm0.02$ & $0.51\pm0.02$ &               &               \\
    & 4.80  & $5.39\pm0.01$  & $1987.38\pm0.01$ &  $4.10\pm0.01$ & $0.81\pm0.02$ &               &               \\
$D$ & 37.0  & $1.56\pm0.04$  & $1988.70\pm0.01$ &  $0.90\pm0.01$ & $0.00\pm0.02$ & --            & --            \\
    & 14.5  & $1.70\pm0.02$  & $1988.80\pm0.01$ &  $1.17\pm0.01$ & $0.10\pm0.02$ &               &               \\
$E$ & 37.0  & $10.12\pm0.04$ & $1991.73\pm0.01$ &  $1.36\pm0.01$ & $0.00\pm0.02$ & $0.76\pm0.14$ & $1.91\pm0.04$ \\
    & 22.0  & $9.72\pm0.07$  & $1991.99\pm0.02$ &  $1.55\pm0.03$ & $0.26\pm0.03$ &               &               \\
    & 14.5  & $7.42\pm0.02$  & $1992.65\pm0.01$ &  $2.12\pm0.01$ & $0.91\pm0.01$ &               &               \\
    & 8.0   & $4.52\pm0.03$  & $1992.88\pm0.01$ &  $1.51\pm0.01$ & $1.15\pm0.01$ &               &               \\
    & 4.8   & $2.15\pm0.01$  & $1993.31\pm0.01$ &  $1.82\pm0.01$ & $1.58\pm0.01$ &               &               \\
$F$ & 37.0  & $5.45\pm0.04$  & $1992.97\pm0.01$ &  $0.99\pm0.01$ & $0.00\pm0.01$ & --            & --            \\
    & 22.0  & $4.08\pm0.06$  & $1993.10\pm0.02$ &  $0.85\pm0.02$ & $0.13\pm0.02$ &               &               \\
$G$ & 37.0  & $1.83\pm0.03$  & $1993.93\pm0.01$ &  $0.80\pm0.01$ & $0.00\pm0.02$ & --            & --            \\
    & 22.0  & $2.46\pm0.03$  & $1993.93\pm0.01$ &  $0.86\pm0.01$ &$-0.01\pm0.02$ &               &               \\
$H$ & 37.0  & $4.36\pm0.03$  & $1995.57\pm0.01$ &  $2.17\pm0.03$ & $0.00\pm0.02$ & $0.13\pm0.07$ & $1.46\pm0.03$ \\
    & 22.0  & $5.30\pm0.03$  & $1995.52\pm0.01$ &  $1.99\pm0.01$ & $0.05\pm0.02$ &               &               \\
    & 14.5  & $4.08\pm0.02$  & $1995.69\pm0.01$ &  $2.43\pm0.01$ & $0.12\pm0.02$ &               &               \\
    & 8.0   & $4.17\pm0.02$  & $1995.61\pm0.03$ &  $4.06\pm0.05$ & $0.04\pm0.04$ &               &               \\
    & 4.8   & $3.51\pm0.02$  & $1995.97\pm0.02$ &  $3.96\pm0.06$ & $0.40\pm0.03$ &               &               \\
$I$ & 22.0  & $1.27\pm0.02$  & $1997.18\pm0.01$ &  $1.13\pm0.01$ & $0.00\pm0.03$ &$-0.32\pm0.19$ & --            \\
    & 14.5  & $1.81\pm0.08$  & $1998.06\pm0.07$ &  $1.40\pm0.04$ & $0.88\pm0.08$ &               &               \\
    & 8.0   & $1.46\pm0.03$  & $1998.69\pm0.01$ &  $1.53\pm0.01$ & $1.51\pm0.03$ &               &               \\
    & 4.8   & $2.41\pm0.01$  & $1999.01\pm0.02$ &  $1.92\pm0.01$ & $1.83\pm0.02$ &               &               \\
$J$ & 37.0  & $3.23\pm0.04$  & $1998.62\pm0.01$ &  $1.96\pm0.01$ & $0.00\pm0.02$ & $0.17\pm0.27$ & $0.40\pm0.01$ \\
    & 22.0  & $3.65\pm0.03$  & $1998.82\pm0.01$ &  $1.70\pm0.02$ & $0.20\pm0.02$ &               &               \\
    & 14.5  & $2.70\pm0.02$  & $1999.32\pm0.01$ &  $1.26\pm0.01$ & $0.70\pm0.02$ &               &               \\
$K$ & 37.0  & $5.51\pm0.05$  & $2000.97\pm0.02$ &  $3.35\pm0.05$ & $0.00\pm0.03$ &$-0.17\pm0.01$ & --            \\
    & 22.0  & $6.26\pm0.06$  & $2001.27\pm0.03$ &  $3.01\pm0.09$ & $0.30\pm0.05$ &               &               \\
    & 14.5  & $6.46\pm0.01$  & $2001.24\pm0.02$ &  $2.51\pm0.01$ & $0.27\pm0.02$ &               &               \\
    & 8.0   & $7.25\pm0.02$  & $2001.16\pm0.01$ &  $3.00\pm0.01$ & $0.19\pm0.02$ &               &               \\
    & 4.8   & $7.93\pm0.01$  & $2001.38\pm0.01$ &  $2.49\pm0.01$ & $0.41\pm0.02$ &               &               \\
\hline  \label{3c345_flares}
\end{tabular}
\end{table*}

\begin{table*}
\begin{center} \caption[Frequency-dependent core shifts]{Comparison of
frequency-dependent core shifts measured from VLBI observations
and calculated from the frequency-dependent time lags for 3C~345.}
\label{3C345_shifts}
\medskip {\footnotesize
\begin{tabular}{llll}
\hline \noalign{\smallskip}

Core shift from VLBI & Core shift from time lags & Epoch & Jet \\
(mas) & (mas) & & comp. \\
\hline \noalign{\smallskip}

$0.05 \pm 0.13$ (10.7 - 5 GHz) & $0.06 \pm 0.02$ (4.8 - 8 GHz) & 1982 & C2 \\
$0.111 \pm 0.007$ (22.2 - 10.7 GHz) & $0.12 \pm 0.02$ (22.2 - 8 GHz) & 1983.5 & C4 \\
$0.05 \pm 0.03$ (5.0 - 8.4 GHz) & $0.09 \pm 0.01$ (4.8 - 8 GHz) & 1993.8 & C7 \\
$0.21 \pm 0.06$ (8.4 - 22.2 GHz) & $0.19 \pm 0.02$ (8.0 - 22.2 GHz) & 1992.8 & C7 \\
$0.33 \pm 0.10$ (5.0 - 22.2 GHz) & $0.28 \pm 0.04$ (4.8 - 22.2 GHz) & 1993.7 & C7 \\
\hline
\end{tabular} }
\end{center}
\end{table*}

\citet*{Biretta_1986} measured core shifts for 3C~345 of $0.05 \pm 0.13$~mas
between 10.7~GHz and 5~GHz in 1982 and $0.111 \pm 0.007$~mas between 22.2~GHz
and 10.7~GHz in 1983.5, during two powerful flares that were associated with
the ejection of the new jet components C2 and C4
(Fig.~\ref{3c345_lcurve}).

Component C2 displayed an apparent speed of
$0.48 \pm 0.02$~mas/yr~\citep{Biretta_1986}; similar speeds were detected
for this component by \citet{Zensus_1995}, $\beta_{app}$ = 0.4--0.53 mas/yr.
The measured time lag between 5~GHz and 8~GHz for the 1982 outburst
corresponding to the ejection of C2 is $0.12 \pm 0.04$ yrs. Following our
method of calculating the core shifts from the time lags as
$\Delta R_{proj}$ [mas] = $\Delta t(\nu)_{obs}$ [yrs] $\cdot {\beta_{app}}$
[mas/yr], this corresponds to a core shift of $0.06 \pm 0.02$ mas. This
``time-lag'' core shift coincides well with the core shift of $0.05 \pm 0.13$
measured directly by aligning VLBI images~\citep{Biretta_1986}.

The jet component C4 displayed an apparent speed of $0.225 \pm
0.015$~mas/yr~\citep{Caproni_2004a}.  \citet{Zensus_1995} and
\citet{Biretta_1986} detected similar speeds of 0.32$\pm$0.15 mas/yr and
0.295$\pm$0.009 mas/yr for this jet component.  The calculated time lag
between 22~GHz and 8~GHz for the 1984 flare is $0.45 \pm 0.08$ yrs. If we
use an average of the three speed estimates, this yields a time-delay core
shift of 0.12$\pm$0.02, which is very close to the core shift of
$0.111 \pm 0.007$ directly measured by~\citet{Biretta_1986}.

\citet{Lobanov_1998} measured core shifts of $\Delta r = 0.05 \pm 0.03$~mas
(5.0--8.4~GHz), $\Delta r = 0.21 \pm 0.06$~mas (8.4~--22.2~GHz), and
$\Delta r = 0.33 \pm 0.10$~mas (5--22.2~GHz) in $\sim 1993$ (see
Table~\ref{3C345_shifts}). We calculated core shifts using our measured
time lags for the flare in 1992: $0.43 \pm 0.01$~yrs (4.8--8.0~GHz),
$0.89 \pm 0.03$~yrs (8.0--22.2~GHz), and $1.32 \pm 0.03$~yrs (4.8--22.0~GHz).
The speed of the jet component C7 ejected during the 1992 flare has been
estimated to be $0.208 \pm 0.025$~mas/yr~\citep{Caproni_2004} and
$0.30 \pm 0.16$~mas/yr~\citep{Ros_2000}. Taking this speed for the newly
%which speed - an average?
ejected jet component C7 and using the measured time lags, we can calculate the
frequency-dependent core shifts, shown in Table~\ref{3C345_shifts}.

Table~\ref{3C345_shifts} clearly shows that the core shifts measured from the
VLBI observations and calculated using our frequency-dependent
time lags and the measured jet-component speeds coincide very well. This
provides a direct test of the proposed method, and suggests
that it can be used to reliably calculate core shifts from total flux-density
light curves. Moreover, this provides direct evidence that the
jet component speed is the actual physical speed of a knot or a
shock, rather than a pattern speed.

\section{Results for the Quasar 3C~345}
\begin{figure}
\centering
\includegraphics[clip,width=8cm]{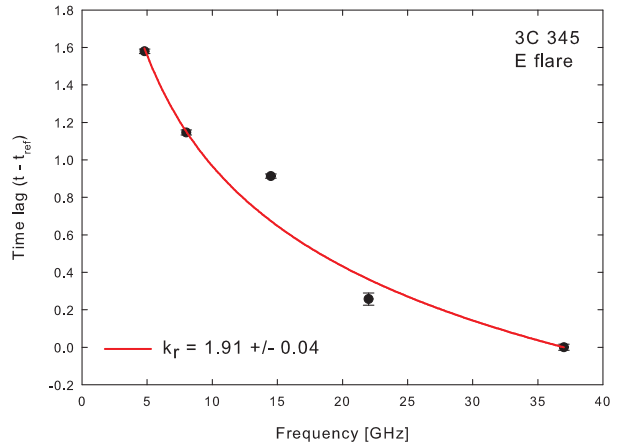} \caption[]{3C~345: Plot of time lag measurements versus frequency, using 37~GHz as the reference
frequency. The solid curve shows the best-fit line $\Delta t = a (\nu^{(-1/k_r)} - 37.0^{(-1/k_r)})$ with a = 5.47$\pm$0.04 and $k_r$ =
1.91$\pm$0.04.} \label{3c345_eflare}
\end{figure}

Looking at Eq. (3), we expect a plot of the time lags versus frequency to
have the form
\begin{equation}
\Delta t(\nu)_{obs} = Constant*(\nu_{a}^{-1/k_{r}} - \nu_{b}^{-1/k_{r}}). \label{eq:k_r2}
\end{equation}
The plots of time lags versus frequency for 3C~345 show that these do, indeed,
follow such a power law, enabling us to fit the time lags and derive $k_{r}$ values from
the fits.  Figure~\ref{3c345_eflare} shows the fit for flare $E$ as an example.
The estimated $k_r$ value from this plot is 1.91$\pm$0.04.
%zzzCheck that this is correct!
We used the
highest frequency, 37~GHz, as the reference frequency for all measured time lags.
Measuring $k_r$ can tell us about the jet geometry, since
the distance from the VLBI core to the base of the jet scales with $k_r$
as $r_{core} \sim \nu^{-1/k_{r}}$ (Eq.~\ref{eq:rcore}). The $k_r$ values
contain information about the distributions of the magnetic field and
electron number density, since $k_r$ depends on the indices $m$ and $n$
(Eq.~\ref{k_r}),
which indicate how the electron number density and magnetic field decrease
along the jet (Eq.~\ref{nm}).  The calculated $k_r$ values for each flare of 3C~345
are listed in the last column of Table~\ref{3c345_flares}. The range of
our $k_r$ values encompasses the average value $k_r = 0.96$ derived
for 3C~345 by~\citet{Lobanov_1998}. We can also use $k_r$ to
calculate the core-region magnetic fields using Eqs.~\ref{eq:omega},
\ref{eq:b1}, and~\ref{eq:bcore}. Our $k_r$ values imply magnetic fields
in the core of 3C~345 $\simeq 0.1$~G.

\begin{figure}
\centering
\includegraphics[clip,width=8cm]{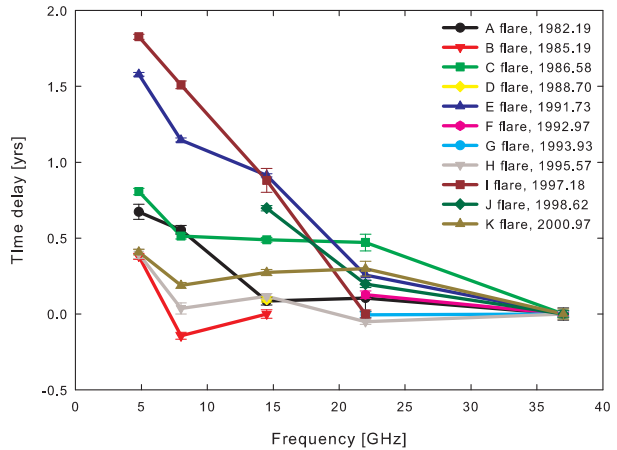} \caption[]{ 3C~345: Time delay of individual events as functions of frequency. } \label{3c345_tdel}
\end{figure}

\begin{figure}
\centering
\includegraphics[clip,width=8cm]{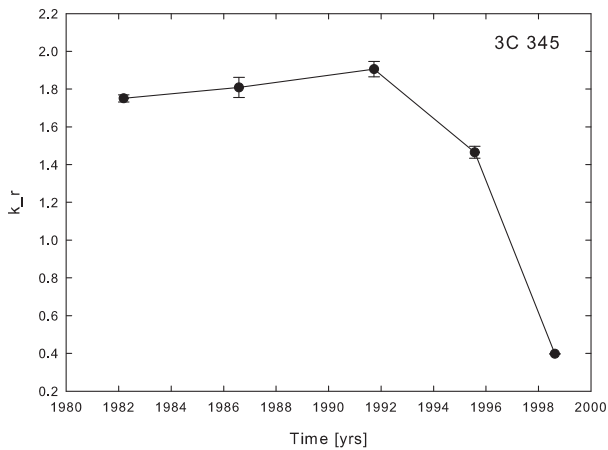} \caption[]{3C~345: Time evolution of $k_{r}$ values. } \label{3c345kr}
\end{figure}

\begin{figure}
\centering
\includegraphics[clip,width=8cm]{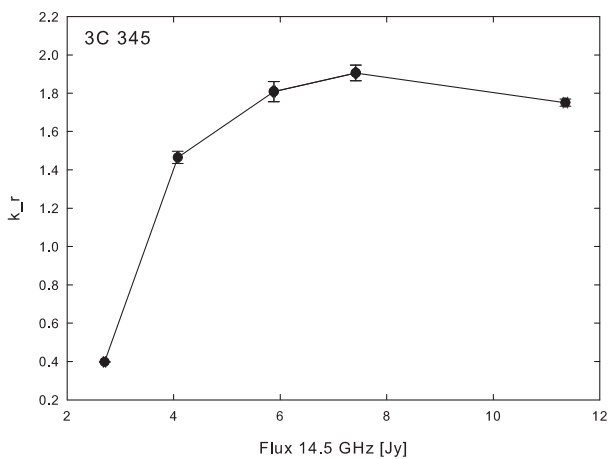} \caption[]{3C~345: correlation between $k_{r}$ values and amplitude of the flares at 14.5 GHz.
The picture shows that the $k_{r}$ values become higher with higher flux
levels.} \label{3c345_kr_flux}
\end{figure}

The measured $k_r$ values for 3C~345 suggest a significant time evolution. The time delays versus frequency are evolving with time.
Figure~\ref{3c345_tdel} shows time lags for individual flares. The maximum time lag is changing between 0.4 yrs and 1.8 yrs.
Figure~\ref{3c345kr} shows that the $k_r$ values are almost at the same level in the period from 1982 to 1992, about 1.8, then begin to
decrease to $k_r \sim 0.4$ in 1999. In this same period, 1982--1990, 3C~345 experienced a major flare, reaching flux-density levels of about 19
Jy (Fig.~\ref{3c345_lcurve}). The lower $k_r$ values seem to correspond to flares with less dramatic amplitudes, suggesting a possible
connection between $k_r$ and the flux-density level. Figure~\ref{3c345_kr_flux} shows $k_r$ versus the flux at 14.5 GHz; there is a clear
tendency for brighter flares to have higher $k_r$ values, reaching a kind of saturation level of about $k_r \sim 1.8$ for fluxes higher than
6~Jy at 14.5~GHz.

%\begin{figure}
%\centering
%\includegraphics[clip,width=8cm]{kr_hist.eps} \caption[]{Distribution of $k_r$ values for 50 flares. } \label{kr_hist}
%\end{figure}

\begin{table*}
\caption{Calculated equipartition magnetic fields and distances between the radio core and the base of the jet.} \centering
\begin{minipage}{110mm}
\begin{tabular}{clllll} \hline

$\nu$   & $\Delta r$ & $\Omega_{r\nu}$ & $r_{core}(\nu)$ & $B_{1pc}$     & $B_{core}(\nu)$ \\
$[GHz]$ & [mas]      & [pc$\cdot$GHz]        & [pc]            & [G]           & [G]             \\
%        &            &                 &                 & ($k_{r} = 1$) & ($k_{r} = 1$)   \\
\hline
\multicolumn{6}{c}{ {\bf1641+399 (3C~345)}: Flare $H$, $k_{r} = 1.46\pm0.03$, $\beta_{app} = 0.37\pm0.03$\footnotemark[1]} \\
4.8/37  & $0.15\pm0.02$   & $3.79\pm0.44$ & $6.79\pm0.78$ & $0.46\pm0.08$ & $0.07\pm0.01$ \\
 8/37   & $0.01\pm0.01$   & $0.59\pm0.55$ & $1.06\pm0.99$ & $0.11\pm0.16$ & $0.11\pm0.18$ \\
14.5/37 & $0.043\pm0.007$ & $3.76\pm0.61$ & $6.73\pm1.10$ & $0.46\pm0.11$ & $0.07\pm0.02$ \\
22/37   & $-0.019\pm0.007$& $3.50\pm1.29$ & $6.26\pm2.31$ & $0.43\pm0.24$ & $0.07\pm0.05$ \\
 \hline  \label{table_mfield}
\end{tabular}
\medskip
Col.(1): frequency; Col.(2): calculated core shift from frequency-dependent time delays; Col.(3): $\Omega_{r\nu}$; Col.(4): distance from radio
core to the base of the jet; Col.(5): equipartition magnetic field at 1 pc; Col.(6): equipartition magnetic field in the core. References:
\footnotemark[1]~\citet{Kellermann_2004}.
\end{minipage}
\end{table*}

\section{Discussion and Conclusion}
\label{discussion}

We have introduced a new method for calculating core shifts from time lags between the maxima of single-dish flares at different frequencies.
The method assumes that the observed velocity of jet components is the actual speed of matter in the jet.

Applying the method to the 3C~345 VLBI jet and integrated light curves from 4.8 to 37 GHz, we found that the directly measured
frequency-dependent core shifts and the core shifts calculated with our new method agree very well, supporting the validity of the method and
our assumptions. In particular, this provides direct evidence that the observed component speeds in the VLBI jet represent the actual physical
speeds of these components, rather than the pattern speed of a shock. This technique should also be checked using more sources, which we plan
to do in a future study.

We have used our method to derive $k_r$ values for 3C~345. We find clear evidence for variability of $k_r$, with the measured values ranging
from 0.4 to 1.9.  We find evidence that $k_r$ increases with the core flux level, reaching saturation at a value of $\simeq 1.8$ above core
fluxes of about 6~Jy. In principle, time evolution of $k_{r}$ could come about due to changes in the core spectral index, magnetic-field
distribution, or electron number-density distribution, since $k_{r}$ depends on $\alpha$, $m$, and $n$ (see formula~\ref{k_r}). There is no
obvious relationship between the value of $k_r$ and the core spectral index (Table~\ref{3c345_flares}). Unfortunately, we cannot directly
separate the $m$ and $n$ values from the $k_r$ equation (the only exception is if $k_r$ is close to unity, i.e., close to equipartition, in
which case it is reasonable to infer $m = 1$ and $n=2$). Therefore, we cannot unambiguously prove that $k_{r}$ variations are due, for example,
purely to variations in $\alpha$, $m$ or $n$; it is likely that all three parameters contribute to time variability of $k_{r}$. {Time
variability of $k_r$ values implies that frequency-dependent core shifts are changing with time, suggesting that in order to match the
astrometric catalogues it is necessary to have simultaneous multi-frequency observations.}

Using our $k_r$ values we can estimate the distance of the radio core from the base of the jet, the equipartition magnetic field at 1 pc
distance, and the equipartition magnetic field in the core (\citet{Lobanov_1998} and~\citet{Hirotani_2005}). Since not all outbursts are in the
equipartition regime as was shown in the previous sections, we have selected for magnetic fields calculations one outburst $H$ with $k_{r}$
value from a Table~\ref{3c345_flares} closest to unity (and therefore to equipartition regime). We have used intrinsic jet half-opening angle
$\theta=0.5^\circ$~\citep{Jorstad_2005}, jet viewing angle $\varphi = 2.7^\circ$~\citep{Jorstad_2005}, Doppler factor $\delta =
7.8$~\citep{Hovatta_2009}, and luminosity distance $D_L = 3473$ Mpc. Table~\ref{table_mfield} shows calculated frequency-dependent core shifts,
distances between the radio core and the base of the jet, magnetic field at 1~pc, and magnetic field in the core for individual pairs of
frequencies. We have not used 8~GHz data in the analysis, since the frequency-dependent time lag for the 8/37 GHz pair of frequencies is not
very reliable. In error analysis we took into account only the errors in time lags measurements and apparent speeds of jet components. The
averaged magnetic field in the core of 3C~345 is $B_{core} = 0.07\pm0.02$ G and magnetic field at 1~pc is $B_{1pc} = 0.45\pm0.09$ G.

By calculating the frequency-dependent core shifts from the frequency-dependent time delays measured for integrated light curves, we can study
the long-term evolution of the core shifts and calculate the core shifts for any particular time when long-term radio light-curves are
available. Since the total flux-density light curves covering more than 30 years are available for dozens of radio sources, this makes it
possible, in principle, to calculate the core-shift evolution over more than 30 years without constructing and aligning VLBI maps at multiple
frequencies. The only input needed from direct VLBI observations is the apparent speeds of jet components (which can be measured from a series
of observations at a single frequency).

%\begin{figure}
%\centering
%\includegraphics[clip,width=8cm]{s_b1.eps} \caption[]{Correlation between total flux-density at 4.8 GHz (corrected for redshift) and magnetic
%field at 1 parsec $B_{1}$.} \label{s_b1}
%\end{figure}

\section*{Acknowledgements}
N. A. Kudryavtseva was supported for this research through a stipend from the International Max Planck Research School (IMPRS) for Radio and
Infrared Astronomy at the Universities of Bonn and Cologne. We would like to thank Shane O'Sullivan for useful discussions. This publication
has emanated from research conducted with the financial support of Science Foundation Ireland. The UMRAO has been supported from the series of
grants from the NSF and NASA and from the University of Michigan.

\label{lastpage}

\end{document}